\documentclass{ws-ijmpd}
\usepackage[super,compress]{cite}
\usepackage{color}
\usepackage{amsmath}	
\begin{document}
\markboth{H. ARAKIDA}
{General Relativistic Aberration and Angle of Light Ray in Kerr Spacetime}

%
\catchline{}{}{}{}{}
%

\title{General Relativistic Aberration Equation and Measurable
Angle of Light Ray in Kerr Spacetime}

\author{Hideyoshi ARAKIDA}

\address{College of Engineering, Nihon University,
Koriyama, Fukushima 9638642 JAPAN\\
arakida.hideyoshi@nihon-u.ac.jp}



\maketitle

\begin{history}
\received{Day Month Year}
\revised{Day Month Year}
\end{history}

\begin{abstract}
We will mainly discuss the measurable angle (local angle) of 
the light ray $\psi_P$ at the position of the observer $P$
instead of the total deflection angle (global angle) $\alpha$
in Kerr spacetime. We will investigate not only the effect of 
the gravito-magnetic field or frame dragging due to the spin 
of the central object but also the contribution of the motion of the observer 
with a coordinate radial velocity $v^r = dr/dt$ and a coordinate 
transverse velocity $bv^{\phi} = bd\phi/dt$ where 
$b \equiv L/E$ is the impact parameter 
($L$ and $E$ are the angular momentum and the energy 
of the light ray, respectively) and $v^{\phi} = d\phi/dt$ is a coordinate 
angular velocity. $v^r$ and $bv^{\phi}$ are computed from the components of 
the 4-velocity of the observer $u^r$ and $u^{\phi}$, respectively. 
Because the motion of observer causes an aberration, we will employ 
the general relativistic aberration equation to obtain the measurable 
angle $\psi_P$ which is determined by the 4-momentum of 
the light ray $k^{\mu}$ and the 4-momentum of the radial null geodesic $w^{\mu}$ 
as well as the 4-velocity of the observer $u^{\mu}$.  
The measurable angle $\psi_P$ given in this paper can be applied not only 
to the case of the observer located in an asymptotically flat region 
but also to the case of the observer placed within the curved and 
finite-distance region. Moreover, when the observer is in radial motion, 
the total deflection angle $\alpha_{\rm radial}$ can be expressed by 
$\alpha_{\rm radial} = (1 + v^r)\alpha_{\rm static}$; this
is consistent with the overall scaling factor $1 - v$ 
instead of $1 - 2v$ with respect to the total deflection angle 
$\alpha_{\rm st atic}$ in the static case 
($v$ is the velocity of the lens object).
On the other hand, when the observer is in transverse motion,
the total deflection angle is given by the form
$\alpha_{\rm transverse} = (1 + bv^{\phi}/2)\alpha_{\rm static}$
if we define the transverse velocity as having the form $bv^{\phi}$.
\end{abstract}


\ccode{PACS numbers:95.30.Sf, 98.62.Sb, 98.80.Es, 04.20.-q, 04.20.Cv}


\section{Introduction\label{sec:intro}}
Since the establishment of the general theory of relativity by
Einstein in 1915-1916, this theory of gravitation has been
subjected to various experimental verifications and the results
have proved the validity of the general theory of relativity. 
Among many experiments, the measurement of the bending of a light 
ray is a historical classical test that was carried out immediately 
after the development of general relativity\cite{dyson_etal1920}. 
In fact, Einstein considered the influence of the gravitational field 
on the path of a light ray and recognized the importance of its 
measurement from an early stage of constructing the general theory 
of relativity\cite{einstein1911,einstein1915}.

Even today, the observation of the light deflection remains important
from the standpoint of the verification of gravitational 
theories\cite{will2014}.
Currently, the measurement accuracy of the VLBI 
(very long baseline Interferometry) is the order of $0.1 \sim 10$ 
mas (milli-arcsecond) or more better and the GAIA mission can determine 
the position of the stars in our galaxy within the order 
$1 \sim 100\ \mu$as (micro-arcsecond). 
Further some space missions have 
been designed to test the gravitational theories such as 
LATOR\cite{turyshev_etal2009} and ASTROD/ASTROD I\cite{ni2008}. 
In particular, the measurement accuracy in an angle of the LATOR mission 
is expected to reach the order of $0.01 ~{\rm picorad} = 10^{-14}~{\rm rad} 
\simeq 2.0 ~{\rm nas}$ (nano-arcsecond); 
see Fig. 3 and Table 1 in\cite{turyshev_etal2009}. 
The light deflection also plays an important role in 
the gravitational lensing which is a powerful tool in the field of 
astrophysics and cosmology, 
see, e.g.,\cite{schneider_etal1999,schneider_etal2006} 
and the references therein.

In general, the celestial objects are rotating and this property can 
be described by the Kerr metric\cite{kerr1963}. The rotation of a
massive object induces a gravito-magnetic field or frame dragging 
which is sometimes called the Lense--Thirring 
effect\cite{thirring1918,lense_thirring1918} characterized 
by the spin parameter $a \equiv J/m$ where $J$ and $m$ are the angular 
momentum and the mass of the celestial object, respectively.
Moreover measuring the spin parameter of a celestial object
as well as its mass leads to testing the no hair (uniqueness)
theorem, e.g.,\cite{carter1971,israel1997,israel1998}.
Then the total deflection angle $\alpha$ in Kerr spacetime has been 
investigated by many authors, and it is found that the correction 
due to the rotation of a celestial object becomes 
$\alpha_{\rm Kerr} = -4am/b^2$ when the observer and the source of 
the light ray are placed in an asymptotically flat region
in which $b \equiv L/E$ is the impact parameter and 
$L$ and $E$ are the angular momentum and the energy of the light ray, 
respectively; see, 
e.g.,\cite{epstein1980,richter_matzner1982,edery_godin2006,sereno_deluca2006,ono_etal2017}.
Furthermore, several authors have discussed the light deflection under
more general conditions and extended the Kerr-type model; 
see, e.g.,\cite{goicoechea_etal1992,kraniotis2005,iyer_hansen2009,sultana2013,kraniotis2014,he_lin2016,he_lin2017a,he_lin2017b,jiang_lin2018,uniyal_etal2018}
and the references therein.

While from a kinematic point of view, several authors considered 
the contribution of the motion of the lens object to the total deflection 
angle in the context of Schwarzschild spacetime. When the lens object is 
moving radially with velocity $v$, the total deflection 
angle $\alpha_{\rm radial}$ is characterized by the overall scaling factor
depending on the radial velocity $v$ 
with respect to the total deflection angle $\alpha_{\rm static}$
in the case of a static lens.
However, a discrepancy exists in the expression of the overall 
scaling factor; $1 - v$ by e.g.\cite{pyne_birkinshaw1993,frittelli2003,wucknitz_sperhake2004},
and $1 - 2v$ by, e.g.,\cite{capozziello_etal1999,sereno2002}.
To solve the geodesic equation of the light ray,
Pyne and Birkinshaw\cite{pyne_birkinshaw1993} developed the method
which is generalization of the Sachs-Wolfe technique\cite{sachs_wolfe1967}.
Frittelli\cite{frittelli2003} simply used the standard aberration formula
derived from the Lorentz transformation and converted the angle 
$\theta$. Wucknitz and Sperhake\cite{wucknitz_sperhake2004} 
calculated the null geodesic equation by using the coordinate 
transformation (the Lorentz transformation.
While, Capozziello {\it et al}\cite{capozziello_etal1999} directly 
integrated the second-order null geodesic equation with 
the vector potential due to the moving mass, and Sereno\cite{sereno2002} 
computed the aberration effect based on the thin lens approximation of 
gravitational lensing and the refraction index which includes the vector 
potential of the mass current.
Currently the contribution of the velocity of the lens object 
to the path of the light ray is observable by VLBI; 
for instance, the effect of the motion of Jupiter on the light ray 
from the quasar J0842+1835, $51 ~\mu{\rm as}$, has been observed 
by VLBA\cite{kopeikin2003}.

In this paper, we will discus mainly on the measurable angle 
(local angle) $\psi_P$ at the position of observer $P$ in Kerr
spacetime.\footnote{We note that in this paper we will not 
investigate deeply the argument involving the total deflection angle 
(global angle) $\alpha$ in curved spacetime.
The reason is that the definition of the total deflection angle 
remains ambiguous because the parallel postulate 
does not hold in a curved spacetime;
the total deflection angle can be obtained only when both the observer 
$R$ and the light source $S$ are placed in asymptotically flat
regions. In this case, due to the parallel postulate, the total deflection 
angle $\alpha$ can be expressed as the twice of the measurable angle 
$\psi_P$, $\alpha = 2\psi_P$ where $P$ means the position of the observer 
$R$. However, in the curved spacetime, we cannot express the total 
deflection angle as $\alpha = 2\psi_P$ because the parallel postulate 
does not hold. 
Although to overcome this situation, some authors considered the
application of the Gauss--Bonnet theorem \cite{ishihara_etal2016,arakida2018}
to define and calculate the total deflection angle $\alpha$ on 
curved spacetime, it has not yet been resolved and further consideration 
is necessary.}
Our aim is to show not only the contribution of the spin parameter 
$a$ of the central object but also the effect of the motion of the observer, 
the coordinate radial velocity $v^r = dr/dt$ and the coordinate transverse
velocity $bv^{\phi} = bd\phi/dt$ ($v^{\phi} = d\phi/dt$ means the coordinate 
angular velocity). Due to the motion of the observer, 
the direction of the light ray coming to the observer 
changes and this effect is known as an aberration. 
To take into the the effect of an aberration and compute 
the measurable angle $\psi_P$, we adopt the general relativistic aberration 
equation\cite{pechenick_etal1983,lebedevlake2013,lebedevlake2016}.
The measurable angle $\psi_P$ is expressed by the tangent vector 
(4-momentum) $k^{\mu}$ of the light ray $\Gamma_k$ 
that we investigate and the tangent vector $w^{\mu}$ along the radial null 
geodesic $\Gamma_w$ connecting the center $O$ and the position of observer 
$P$ as well as the 4-velocity $u^{\mu}$ of the observer. 
The coordinate velocities $v^r$ and $v^{\phi}$ are related to the components 
of the 4-velocity of the observer $u^r$ and $u^{\phi}$, respectively.
The general relativistic aberration equation enables us to compute 
the contribution of the velocity effect to the bending angle of 
a light ray more easily and straightforwardly because the null geodesic 
of light ray $\Gamma_k$ and the radial null geodesic $\Gamma_w$ 
are independent of the velocity effect. 
The measurable angle $\psi_P$ calculated below is not limited 
to the case of the observer located in an asymptotically flat 
region of spacetime; it can also be 
applied to the case of the observer placed within the curved region.

This paper is organized as follows: in section \ref{sec:trajectory},
the trajectory of light ray in Kerr spacetime is derived from the
first-order differential equation of the null geodesic.
In section \ref{sec:aberration}, the general relativistic
aberration equation is introduced, and in section
\ref{sec:angle} the measurable angle $\psi_P$ is calculated for
the case of the static observer, the observer in radial motion
and the observer in transverse motion. Finally, section
\ref{sec:conclusions} is devoted to presenting conclusions.
\section{Light Trajectory in Kerr Spacetime\label{sec:trajectory}}
Kerr spacetime in Boyer--Lindquist coordinates
$(t, r, \theta, \phi)$ can be rearranged as \cite{bl1967} 
\begin{eqnarray}
 ds^2 &=& g_{\mu\nu}dx^{\mu}dx^{\nu}
  \nonumber\\
  &=& - \left(1 - \frac{2mr}{\Sigma}\right)dt^2
  + \frac{\Sigma}{\Delta}dr^2  
  - \frac{4mar\sin^2\theta}{\Sigma}dtd\phi
  \nonumber\\
  &+& \Sigma d\theta^2
  + \left(r^2 + a^2 + \frac{2ma^2r\sin^2\theta}{\Sigma}\right)
  \sin^2\theta d\phi^2,
  \label{eq:kerr-metric1}\\
 \Sigma &=& r^2 + a^2\cos^2\theta,\quad
  \Delta = r^2 + a^2 - 2mr,
  \label{eq:sigma_delta}
\end{eqnarray}
where the Greek indices such as $\mu, \nu$ run from 0 to 3,
$m$ is the mass of the central object, $a \equiv J/m$ is a spin
parameter ($J$ is the angular momentum of the central object). 
Throughout this paper we use the geometrical unit $c = G = 1$.

For the sake of brevity, we take the equatorial plane
$(\theta = \pi/2,\ d\theta = 0)$ as an orbital plane of the
light ray, then the line element becomes
\begin{eqnarray}
 ds^2 = -A(r)dt^2 + B(r)dr^2 + 2C(r)dtd\phi + D(r)d\phi^2,
  \label{eq:kerr-metric3}
\end{eqnarray}
where $A(r), B(r), C(r)$, and $D(r)$ are
\begin{eqnarray}
 A(r) &=& 1 - \frac{2m}{r},
  \label{eq:Ar}\\
 B(r) &=& \left(1 - \frac{2m}{r} + \frac{a^2}{r^2}\right)^{-1},
  \label{eq:Br}\\
 C(r) &=& - \frac{2ma}{r},
  \label{eq:Cr}\\
  D(r) &=& r^2 + a^2 + \frac{2ma^2}{r}.
  \label{eq:Dr}
\end{eqnarray}
We mention that it may be a restriction in theoretical model construction 
to assume that the trajectory of light is on the equatorial plane of the
Sun if actual observations in solar system are taken into account.
However, the planned missions such as LATOR and ASTROD/ASTROD I aim to test 
the theory of gravity by setting three laser baselines over the equatorial 
plane of the Sun using multiple spacecrafts. 
Therefore, it is reasonable to take photon orbits on the equatorial 
plane of the Sun when discussing the observability in future missions.
In addition, the speed of spacecrafts and the Earth relative to the speed of 
light as well as the value of the angular momentum of the Sun are sufficiently
small. Therefore, in this paper, we discuss the observability imposing 
the slow motion and slow rotation approximation.

From the relation of two constants of motion, the energy $E$ and the
angular momentum $L$, 
$dt/d\lambda$ and $d\phi/d\lambda$ are expressed as
\begin{eqnarray}
 \frac{dt}{d\lambda}
  &=&
  \frac{ED(r) + LC(r)}{A(r)D(r) + C^2(r)},
  \label{eq:dtdl}\\
 \frac{d\phi}{d\lambda}
  &=&
  \frac{LA(r) - EC(r)}{A(r)D(r) + C^2(r)},
  \label{eq:dpdl}
\end{eqnarray}
in which $\lambda$ is an affine parameter.
From the null condition $ds^2 = 0$ and Eqs. (\ref{eq:kerr-metric3}),
(\ref{eq:dtdl}), and (\ref{eq:dpdl}), the geodesic equation
of a light ray is expressed as
\begin{eqnarray}
 \left(\frac{dr}{d\phi}\right)^2
  =
  \frac{A(r)D(r) + C^2(r)}{B(r)[bA(r) - C(r)]^2}
  [- b^2A(r) + 2bC(r) + D(r)],
  \label{eq:geodesiceq1}
\end{eqnarray}
where we introduced another constant, the impact parameter $b$ as
\begin{eqnarray}
 b \equiv \frac{L}{E}.
  \label{eq:impact}
\end{eqnarray}
Although Eq. (\ref{eq:geodesiceq1}) gives the exact expression
of the null geodesic equation,
it is too complicated to yield the trajectory of a light ray. 
Thus, let us introduce a small dimensionless expansion 
parameter $\varepsilon$ which is $\varepsilon = m/b$ and $a/b$, 
and expand Eq. (\ref{eq:geodesiceq1}) up to the order
${\cal O}(\varepsilon^2)$;
\begin{eqnarray}
 \left(\frac{du}{d\phi}\right)^2
  &=&
  \frac{1}{b^2} - u^2 - 2a^2u^4 + \frac{3a^2u^2}{b^2}
  + 2mu^3 - \frac{4mau}{b^3}
  + {\cal O}(\varepsilon^3).
  \label{eq:geodesiceq4}
\end{eqnarray}
in which we changed the variable $r$ by $u = 1/r$ and 
inserted Eqs. (\ref{eq:Ar}), (\ref{eq:Br}), (\ref{eq:Cr}),
and (\ref{eq:Dr}) into Eq. (\ref{eq:geodesiceq1}).
Throughout this paper, we use $\varepsilon$ to represent 
the order of the approximation and the expressions such as 
${\cal O}(\varepsilon^2)$ and ${\cal O}(\varepsilon^3)$ are the simplified 
notation meaning the combination of $m/b$ and $a/b$.
According to the standard perturbation scheme, let the solution 
$u = u(\phi)$ be
\begin{eqnarray}
 u = \frac{\sin \phi}{b} + \varepsilon u_1 + \varepsilon^2 u_2,
  \label{eq:trajectory1}
\end{eqnarray}
where $u_0 = \sin\phi/b$ is the zero-th order solution, and
$\varepsilon u_1$ and $\varepsilon^2 u_2$ are the first 
order ${\cal O}(\varepsilon)$ and second order ${\cal O}(\varepsilon^2)$ 
corrections to $u_0$, respectively. Substituting 
Eq. (\ref{eq:trajectory1}) into Eq. (\ref{eq:geodesiceq4}) 
and collecting the same order terms, the equation describing the light 
trajectory is given by up to the order ${\cal O}(\varepsilon^2)$:
\begin{eqnarray}
\frac{1}{r}
  &=& \frac{\sin\phi}{b}
  + \frac{m}{2b^2}(3 + \cos 2\phi)\nonumber\\
 &+& \frac{1}{16b^3}
  \left\{
   m^2\left[
       37\sin\phi + 30(\pi - 2\phi)\cos\phi - 3\sin 3\phi
      \right]
   + 8a^2\sin^3\phi - 32am
  \right\}\nonumber\\
 &+& {\cal O}(\varepsilon^3),
  \label{eq:trajectory2}
\end{eqnarray}
here the integration constant is chosen so as to maximize $u$ 
(minimize $r$) at $\phi = \pi/2$. 
\section{General Relativistic Aberration Equation
 \label{sec:aberration}}
\begin{figure}[htbp]
\begin{center}
 \includegraphics[scale=0.2,clip]{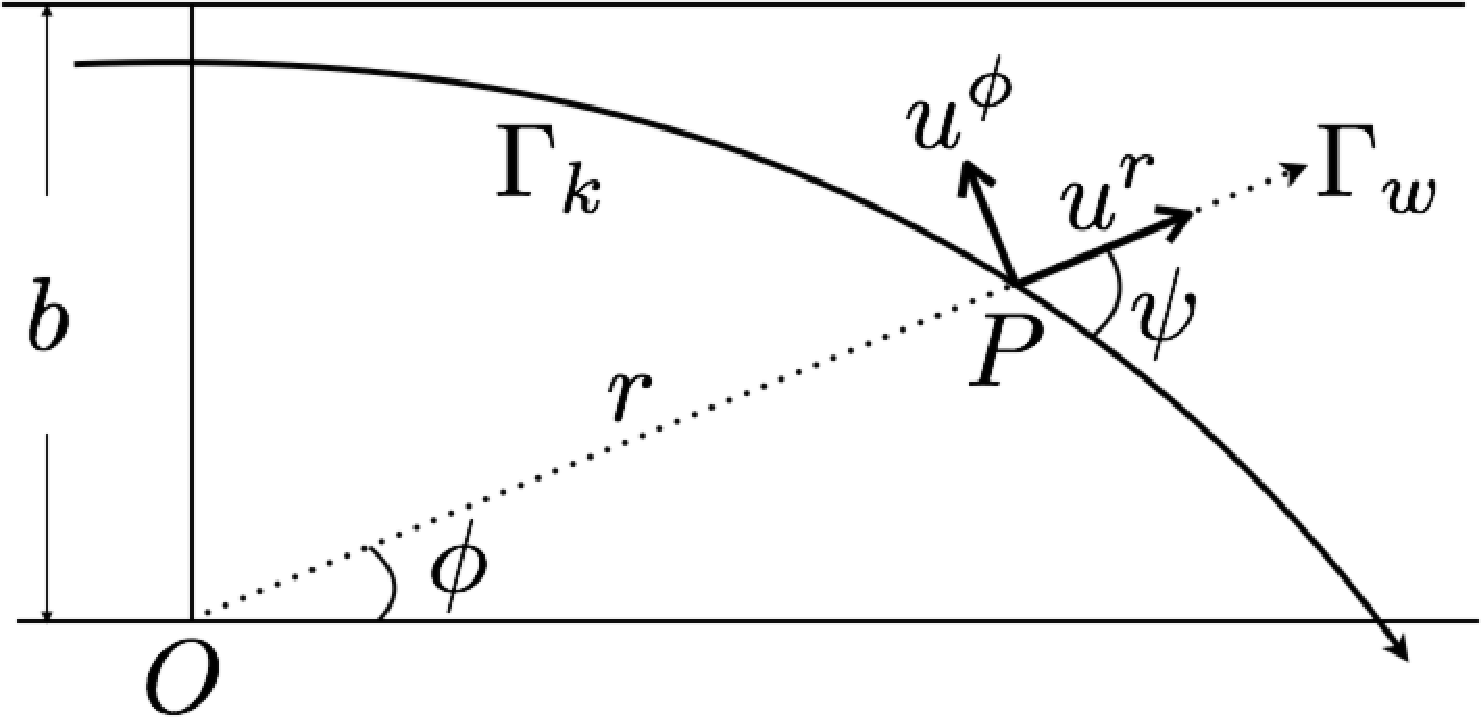}
 \caption{Schematic diagram of light trajectory.
 $\Gamma_k$ (bold line) is the trajectory of the light ray
 which we now investigate, $\Gamma_w$ (dotted line) is the radial null
 geodesic connecting the center $O$ and the position of observer $P$
 and the measurable angle $\psi$ is the angle of intersection between
 $\Gamma_k$ and $\Gamma_w$ at $P$. Two bold vectors $u^r$ and $u^{\phi}$
 at $P$ indicate the directions of the $r$ (radial) and $\phi$ (transverse)
 components of the 4-velocity $u^{\mu}$. The direction of the time component
 $u^t$ is perpendicular to this schematic plane.
 \label{fig:arakida-fig1}}
\end{center}
\end{figure}

In this section, we briefly summarize the outline of the derivation of 
the general relativistic aberration equation discussed 
in\cite{pechenick_etal1983,lebedevlake2013,lebedevlake2016}.
Let $k^{\mu}$ be the 4-momentum (the tangent vector) of 
the light ray $\Gamma_k$ which we now investigate and $w^{\mu}$ be 
the 4-momentum of the radial null geodesic $\Gamma_w$ connecting
the center $O$ and the position of observer $P$, and 
the 4-velocity of the observer is $u^{\mu} = dx^{\mu}/d\tau$ 
($\tau$ is the proper time of the observer).
See Figure \ref{fig:arakida-fig1} for the schematic diagram of
the light trajectory.
Here, we divide the null vectors $k^{\mu}$ and $w^{\mu}$ into the 
two parts,
\begin{align}
k^{\mu} &= k_{\parallel}^{\mu} + k_{\perp}^{\mu},\\
w^{\mu} &= w_{\parallel}^{\mu} + w_{\perp}^{\mu}, 
\end{align}
where $k_{\parallel}^{\mu}$ and $w_{\parallel}^{\mu}$ are the 
components of $k^{\mu}$ and $w^{\mu}$ being parallel to the 4-velocity 
$u^{\mu}$; $k_{\perp}^{\mu}$ and $w_{\perp}^{\mu}$ 
are the projections of $k^{\mu}$ and $w^{\mu}$ onto the space (3-surface) 
of the observer being orthogonal to $u^{\mu}$.
Here we introduce the projection tensor $P_{\mu\nu}$ as
\begin{align}
 P^{\mu}_{\nu} &= \delta^{\mu}_{\nu} + u^{\mu}u_{\nu},\\
 P_{\mu\nu} &= g_{\mu\lambda}P^{\lambda}_{\nu} = 
 g_{\mu\nu} + u_{\mu}u_{\nu},
 \label{eq:projection1}
\end{align}
in which $\delta^{\mu}_{\nu}$ is the Kronecker's delta symbol.
Then $k_{\perp}^{\mu}$ and $w_{\perp}^{\mu}$ are expressed as,
\begin{align}
 k_{\perp}^{\mu} 
 &= P^{\mu}_{\nu}k^{\nu} 
 = (\delta^{\mu}_{\nu} + u^{\mu}u_{\nu})k^{\nu},
 \label{eq:projection2}
 \\
 w_{\perp}^{\mu} 
 &= P^{\mu}_{\nu}w^{\nu}
 = (\delta^{\mu}_{\nu} + u^{\mu}u_{\nu})w^{\nu}.
 \label{eq:projection3}
\end{align}
The measurable angle $\psi_P$ is defined as the intersection 
angle between two vectors $k_{\perp}^{\mu}$ and $w_{\perp}^{\mu}$ 
in the local frame of the observer where the local metric $\eta_{\mu\nu}$
of the observer frame should be consistent with $g_{\mu\nu}$.
Then,
\begin{align}
 \eta_{\mu\nu}k_{\perp}^{\mu}w_{\perp}^{\nu}
 =
 g_{\mu\nu}k_{\perp}^{\mu}w_{\perp}^{\nu}
 =
 g_{\mu\nu}P^{\mu}_{\lambda}k^{\lambda}P^{\nu}_{\delta}k^{\delta}
 =
 P_{\lambda\nu}k^{\lambda}P^{\nu}_{\delta}k^{\delta}.
\end{align}
Further, it is easy to check that 
\begin{align}
 P_{\lambda\nu}P^{\nu}_{\delta} = P_{\lambda\delta}.
 \label{eq:projection4}
\end{align}
Therefore, it is found,
\begin{align}
 \eta_{\mu\nu} = P_{\mu\nu}.
 \label{eq:projection5}
\end{align}
By using Eqs. (\ref{eq:projection1}), (\ref{eq:projection2}), 
(\ref{eq:projection3}), (\ref{eq:projection4}), and (\ref{eq:projection5}),
the measurable angle $\psi_P$ can be obtained by following 
invariant cosine formula;
\begin{align}
 \cos \psi_P 
 &= \frac{\eta_{\mu\nu}k_{\perp}^{\mu}w_{\perp}^{\nu}}
{\sqrt{\eta_{\mu\nu}k_{\perp}^{\mu}k_{\perp}^{\nu}}
 \sqrt{\eta_{\mu\nu}w_{\perp}^{\mu}w_{\perp}^{\nu}}}
 \nonumber\\
 &=
 \frac{g_{\mu\nu}k^{\mu}w^{\nu} + u_{\mu}u_{\nu}k^{\mu}w^{\nu}}
 {\sqrt{g_{\mu\nu}k^{\mu}k^{\nu} + u_{\mu}u_{\nu}k^{\mu}k^{\nu}}
 \sqrt{g_{\mu\nu}w^{\mu}w^{\nu} + u_{\mu}u_{\nu}w^{\mu}w^{\nu}}}.
  \label{eq:aberration1}
\end{align}
Because $k^{\mu}$ and $w^{\mu}$ are the null vectors,
$g_{\mu\nu}k^{\mu}k^{\nu} = 0$ and 
$g_{\mu\nu}w^{\mu}w^{\nu} = 0$, Eq. (\ref{eq:aberration1}) is written as,
\begin{eqnarray}
 \cos\psi_P =
  \frac{g_{\mu\nu}k^{\mu}w^{\nu}}
  {(g_{\mu\nu}u^{\mu}k^{\nu})(g_{\mu\nu}u^{\mu}w^{\nu})} + 1.
  \label{eq:aberration2}
\end{eqnarray}
Because Eq. (\ref{eq:aberration2}) includes the 4-velocity of 
the observer $u^{\mu}$, it enables us to calculate the influence of 
the motion of the observer on the measurable angle $\psi_P$. 
\section{Measurable Angle of Light Ray\label{sec:angle}}
We are working in the equatorial plane $\theta = \pi/2$,
then the components of $k^{\mu}$ and $w^{\mu}$ are
\begin{eqnarray}
 k^{\mu} = (k^t, k^r, 0, k^{\phi}),\quad
 w^{\mu} = (w^t, w^r, 0, 0).
\end{eqnarray}
From the null condition, $g_{\mu\nu}k^{\mu}k^{\nu} = 0$ and
$g_{\mu\nu}w^{\mu}w^{\nu} = 0$, $k^t$ and $w^t$ are given by
\begin{eqnarray}
 k^t &=& \frac{C(r)k^{\phi} +
  \sqrt{[C(r)k^{\phi}]^2 + A(r)[B(r)(k^r)^2 + D(r)(k^{\phi})^2]}}
  {A(r)},
  \label{eq:kt}\\
 w^t &=& \sqrt{\frac{B(r)}{A(r)}}w^r,
  \label{eq:wt}
\end{eqnarray}
where we chose the sign of $k^t$ and $w^t$ to be positive.
To calculate the measurable angle $\psi_P$ below, 
we use he following relation:
\begin{eqnarray}
 \frac{k^r}{k^{\phi}}
  = \frac{dr/d\lambda}{d\phi/d\lambda}
  = \frac{dr}{d\phi}.
  \label{eq:krkp}
\end{eqnarray}

\subsection{Measurable Angle by Static Observer}
In the case of a static observer, the component of the 4-velocity of 
the observer, $u^{\mu}$, becomes 
\begin{eqnarray}
 u^{\mu} = (u^t, 0, 0, 0),
\end{eqnarray}
and from the condition $g_{\mu\nu}u^{\mu}u^{\nu} = -1$,
$u^t$ is expressed as
\begin{eqnarray}
 u^t = \frac{1}{\sqrt{A(r)}},
  \label{eq:ut_sta}
\end{eqnarray}
where we assume $u^t$ to be positive. 
Inserting Eqs. (\ref{eq:geodesiceq1}), (\ref{eq:kt}),
(\ref{eq:wt}), (\ref{eq:krkp}), and (\ref{eq:ut_sta})
into Eq. (\ref{eq:aberration2}), we have
\begin{eqnarray}
 \cos\psi_{\rm static}
  = \sqrt{\frac{A(r)[-b^2A(r) + 2bC(r) + D(r)]}{A(r)D(r) + C^2(r)}}.
  \label{eq:cos_aberration_sta1}
\end{eqnarray}
Further, substituting Eqs. (\ref{eq:Ar}), (\ref{eq:Br}), (\ref{eq:Cr}),
(\ref{eq:Dr}), and (\ref{eq:trajectory2})
into Eq. (\ref{eq:cos_aberration_sta1}), and expanding up to the
order ${\cal O}(\varepsilon^2)$, 
the measurable angle $\psi_{\rm static}$ observed by the static observer 
becomes for the range $0 \le \phi \le \pi/2$,
\begin{eqnarray}
 \psi_{\rm static}
  &=& \phi + \frac{2m}{b}\cos\phi
  \nonumber\\
  &+& \frac{1}{8b^2}
  \left\{
   m^2\left[15(\pi - 2\phi) - \sin 2\phi\right]
   - 16ma \cos \phi
	  \right\}
  + {\cal O}(\varepsilon^3).
  \label{eq:psi_sta1}
\end{eqnarray}
and for the range $\pi/2 \le \phi \le \pi$,
\begin{eqnarray}
 \psi_{\rm static}
  &=& \arccos(-\cos\phi) - \frac{2m}{b}\cos\phi
  \nonumber\\
  &-& \frac{1}{8b^2}
  \left\{
   m^2\left[15(\pi - 2\phi) - \sin 2\phi\right]
   - 16ma \cos \phi
	  \right\}
  + {\cal O}(\varepsilon^3).
  \label{eq:psi_sta2}
\end{eqnarray}
Note here that we have divided the expression of $\psi_{\rm static}$
into two cases, Eqs. (\ref{eq:psi_sta1}) and (\ref{eq:psi_sta2}).
This was done in order to utilize the trigonometric identities
$\sqrt{1 - \sin^2\phi} = \cos\phi$ for $0 \le \phi \le \pi/2$ and
$\sqrt{1 - \sin^2\phi} = -\cos\phi$ for $\pi/2 \le \phi \le \pi$.
Henceforth we employ a similar procedure when calculating the 
angle measured by the observer in radial motion, $\psi_{\rm radial}$ and
in transverse motion, $\psi_{\rm transverse}$. Although this
procedure may not be necessary for computing $\psi_{\rm static}$
and $\psi_{\rm radial}$, this treatment is required when
calculating $\psi_{\rm transverse}$; see Eq. (\ref{eq:psi_circ1})
and observe the case of $\phi \rightarrow \pi$.

If the observer and the source of the light ray are located
in an asymptotically flat region at infinity,
$\phi \rightarrow 0$ and $\phi \rightarrow \pi$, respectively,
it is found that
\begin{eqnarray}
 \psi_{\rm static}(\phi \rightarrow 0)
  &=&
  \psi_{\rm static}(\phi \rightarrow \pi)
  \nonumber\\
  &\rightarrow&
  \frac{2m}{b} + \frac{15\pi m^2}{8b^2} - \frac{2ma}{b^2}
  + {\cal O}(\varepsilon^3).
  \label{eq:psi_sta3}
\end{eqnarray}
In this case, the sum of
$\psi_{\rm static}(\phi \rightarrow 0)$ and 
$\psi_{\rm static}(\phi \rightarrow \pi)$ can be
regarded as the total deflection angle $\alpha_{\rm static}$:
\begin{eqnarray}
 \alpha_{\rm static}
  &=&
  \psi_{\rm static}(\phi \rightarrow 0)
  +
  \psi_{\rm static}(\phi \rightarrow \pi)
  \nonumber\\
  &=& \frac{4m}{b} + \frac{15\pi m^2}{4b^2} - \frac{4ma}{b^2}
  + {\cal O}(\varepsilon^3),
  \label{eq:alpah_sta}
\end{eqnarray}
where the third term is the correction due to the rotation of
the central object and it is in agreement with the result
previously obtained by, 
e.g.,\cite{epstein1980,richter_matzner1982,edery_godin2006,sereno_deluca2006,ono_etal2017}.
\subsection{Measurable Angle by Observer in Radial Motion}
In the case of a radially moving observer, the component of
the 4-velocity $u^{\mu}$ is
\begin{eqnarray}
 u^{\mu} = (u^t, u^r, 0, 0),
\end{eqnarray}
and the condition $g_{\mu\nu}u^{\mu}u^{\nu} = -1$ gives the
relation between $u^t$ and $u^r$ as
\begin{eqnarray}
 u^t = \sqrt{\frac{B(r)(u^r)^2 + 1}{A(r)}},
  \label{eq:ut_rad}
\end{eqnarray}
where as in Eq. (\ref{eq:ut_sta}), we take $u^t$ to be a positive value.
Here, let us introduce the coordinate radial velocity $v^r$ as
\begin{eqnarray}
 v^r = \frac{dr}{dt} = \frac{dr/d\tau}{dt/d\tau} = \frac{u^r}{u^t}.
\label{eq:vr1}
\end{eqnarray}
Substituting Eq. (\ref{eq:ut_rad}) into Eq. (\ref{eq:vr1}),
we find
\begin{eqnarray}
 u^r = \frac{v^r}{\sqrt{A(r) - B(r)(v^r)^2}}.
  \label{eq:vr2}
\end{eqnarray}
To compute $\psi_{\rm radial}$ up to the order
${\cal O}(\varepsilon^2, \varepsilon^2v^r)$,
let us impose the slow motion approximation for the velocity of
the observer $v^r \ll 1$. In fact, velocity 
of the observer satisfies this condition; in the case of the spacecraft,
its velocity with respect to the sun is 
$v \approx v_3 \simeq 4.2 \times 10^4 ~{\rm m/s} \ll c$ 
($v_3$ is the third escape velocity with respect to the sun and 
$c$ is the speed of light in vacuum).

We employ the same procedure used to obtain Eqs. (\ref{eq:psi_sta1}) 
and (\ref{eq:psi_sta2}), 
then $\psi_{\rm radial}$ is for the range $0 \le \phi \le \pi/2$,
\begin{eqnarray}
 \psi_{\rm radial}
  &=&
  \phi + v^r\sin\phi
  + \frac{2m}{b}(\cos\phi + v^r)
  \nonumber\\
  &+& \frac{1}{8b^2}
  \left(
   m^2[15(\pi - 2\phi) - \sin 2\phi] - 16am\cos\phi
   \right)\nonumber\\
 &+& \frac{v^r}{16b^2}\left\{
	m^2\left[
      30 (\pi - 2\phi) \cos\phi + 95\sin\phi - \sin 3\phi
     \right]\right.
	\nonumber\\
    &-&\left.16am(1 + \cos 2\phi) - 2a^2(3\sin\phi - \sin 3\phi)\right\}
    + {\cal O}(\varepsilon^3, (v^r)^2, \varepsilon (v^r)^2),
\label{eq:psi_rad1}
\end{eqnarray}
and for the range $\pi/2 \le \phi \le \pi$,
\begin{eqnarray}
 \psi_{\rm radial}
  &=&
  \arccos(-\cos\phi) + v^r\sin\phi +
  \frac{2m}{b}\left(-\cos\phi + v^r\right)
  \nonumber\\
  &-& \frac{1}{8b^2}
  \left(
   m^2[15(\pi - 2\phi) - \sin 2\phi] - 16am\cos\phi
     \right)\nonumber\\
 &+&
  \frac{v^r}{16b^2}
  \left\{
   m^2\left[
   30(\pi - 2\phi) \cos\phi + 95\sin\phi - \sin 3\phi
 \right]\right.
  \nonumber\\
 &-&\left. 16am(1 + \cos 2\phi) - 2a^2(3\sin\phi - \sin 3\phi)\right\}
  + {\cal O}(\varepsilon^3, (v^r)^2, \varepsilon (v^r)^2).
  \label{eq:psi_rad2}
\end{eqnarray}

When the observer and the source of a light ray are located
in an asymptotically flat region at infinity, $\phi \rightarrow 0$
and $\phi \rightarrow \pi$, respectively, and thus 
\begin{eqnarray}
 \psi_{\rm radial}(\phi \rightarrow 0)
  &=& \psi_{\rm radial}(\phi \rightarrow \pi)
  \nonumber\\
  &\rightarrow&
  \left(
  \frac{2m}{b}
  + \frac{15\pi m^2}{8b^2} - \frac{2am}{b^2}
  \right)
  (1 + v^r)
  + {\cal O}(\varepsilon^3, (v^r)^2, \varepsilon (v^r)^2),
  \label{eq:psi_rad3}
\end{eqnarray}
In this case, the total deflection angle $\alpha_{\rm radial}$ becomes
\begin{eqnarray}
 \alpha_{\rm radial}
  &=&
   \psi_{\rm radial}(\phi \rightarrow 0)
   + \psi_{\rm radial}(\phi \rightarrow \pi)
   \nonumber\\
 &=& \left(
  \frac{4m}{b}
  + \frac{15\pi m^2}{4b^2} - \frac{4am}{b^2}
  \right)
  (1 + v^r)
 + {\cal O}(\varepsilon^3, (v^r)^2, \varepsilon (v^r)^2).
  \label{eq:alpha_rad}
\end{eqnarray}
From Eq. (\ref{eq:alpha_rad}), it is found that the total deflection
angle becomes larger than that of the static case when the observer moves
radially away from the center, $v^r > 0$, whereas it becomes smaller
when the observer radially approaches the center, $v^r < 0$.

It is clear from Eqs. (\ref{eq:alpah_sta}) and (\ref{eq:alpha_rad})
that the relation between $\alpha_{\rm static}$ and 
$\alpha_{\rm radial}$ is expressed as
\begin{eqnarray}
 \alpha_{\rm radial} = (1 + v^r)\alpha_{\rm static}.
  \label{eq:alpha_sta_rad}
\end{eqnarray}
Eq. (\ref{eq:alpha_sta_rad}) is in agreement with the result, e.g.,
\cite{pyne_birkinshaw1993,frittelli2003,wucknitz_sperhake2004},
in which the form of the overall scaling factor is $1 - v$
instead of $1 - 2v$ reported by, e.g.,\cite{capozziello_etal1999,sereno2002}.
It should be mentioned that the radial velocity $v$ used in
\cite{pyne_birkinshaw1993,capozziello_etal1999,sereno2002,frittelli2003,wucknitz_sperhake2004}
is that of the lens object while the radial velocity $v^r$ in our case
is that of the observer. 
Then $v^r$ and $v$ can be connected as follows:
\begin{eqnarray}
 v^r = -v.
\end{eqnarray}
\subsection{Measurable Angle by Observer in Transverse Motion}
Let us investigate the case of the observer in transverse motion
which is the motion in a direction perpendicular
to the radial direction in the orbital plane.

The component of the 4-velocity of the observer $u^{\mu}$ is
\begin{eqnarray}
 u^{\mu} = (u^t, 0, 0, u^{\phi}),
\end{eqnarray}
and the condition $g_{\mu\nu}u^{\mu}u^{\nu} = -1$ gives
\begin{eqnarray}
 u^t = \frac{C(r)u^{\phi} +
  \sqrt{[C(r)u^{\phi}]^2 + A(r)[D(r)(u^{\phi})^2 + 1]}}{A(r)},
  \label{eq:ut_circ}
\end{eqnarray}
in which we choose the positive sign for $u^t$.
The coordinate angular velocity $v^{\phi}$ is determined by
\begin{eqnarray}
 v^{\phi} = \frac{d\phi}{dt} = \frac{d\phi/d\tau}{dt/d\tau}
  = \frac{u^{\phi}}{u^t},
  \label{eq:vp1}
\end{eqnarray}
and using Eq. (\ref{eq:ut_circ}), $u^{\phi}$ is obtained by
means of $v^{\phi}$ as
\begin{eqnarray}
 u^{\phi} = \frac{v^{\phi}}
  {\sqrt{A(r) - 2C(r)v^{\phi} - D(r)(v^{\phi})^2}}.
  \label{eq:vp2}
\end{eqnarray}
%
Because $v^{\phi} = d\phi/dt$ is the coordinate angular velocity,
we regard $bv^{\phi}$ as the coordinate transverse velocity
and impose the slow motion approximation $bv^{\phi} \ll 1$ 
as in the case of radial motion.
Similar to the derivation of Eqs. (\ref{eq:psi_rad1}) and 
(\ref{eq:psi_rad2}), $\psi_{\rm transverse}$ is given by up to the order
${\cal O}(\varepsilon^2, \varepsilon^2 bv^{\phi})$
for $0 \le \phi \le \pi/2$,
\begin{eqnarray}
 \psi_{\rm transverse}
  &=&
  \phi + bv^{\phi}\tan\frac{\phi}{2}
  + \frac{2m}{b}\cos\phi\left(1 + \frac{bv^{\phi}}{1 + \cos\phi}\right)
  \nonumber\\
 &+&
  \frac{1}{8b^2}
  \left\{
   m^2[15(\pi - 2\phi) - \sin 2\phi] - 16am\cos\phi
   \right\}\nonumber\\
 &+&\frac{bv^{\phi}}{8b^2(1 + \cos\phi)}
  \Biggl\{
  m^2\left[15(\pi - 2\phi)
      - 16\sin \phi + 7\sin 2\phi + 16\tan \frac{\phi}{2}
     \right]\Biggr.
  \nonumber\\
 &+&
  \Biggl.
   8ma(1 - 2\cos\phi - \cos 2\phi)
  \Biggr\}
  + {\cal O}(\varepsilon^3, (bv^{\phi})^2, \varepsilon (bv^{\phi})^2),
  \label{eq:psi_circ1}
\end{eqnarray}
and for $\pi/2 \le \phi \le \pi$:
\begin{eqnarray}
 \psi_{\rm transverse}
  &=&
  \arccos(-\cos\phi) + bv^{\phi}\cot\frac{\phi}{2}
  - \frac{2m}{b}\cos\phi\left(1 + \frac{bv^{\phi}}{1 - \cos\phi}\right)
  \nonumber\\
 &-& \frac{1}{8b^2}\{m^2[15(\pi - 2\phi) - \sin 2\phi] - 16am\cos\phi\}
  \nonumber\\
 &+& \frac{bv^{\phi}}{8b^2(1 - \cos\phi)}
  \Biggl\{
   m^2\left[
       -15(\pi - 2\phi)
       - 16\sin\phi - 7\sin 2\phi + 16\cot\frac{\phi}{2}
      \right]\Biggr.
   \nonumber\\
 &+& \Biggl. 8am(1 + 2\cos\phi - \cos 2\phi)
  \Biggr\}
  + {\cal O}(\varepsilon^3, (bv^{\phi})^2, \varepsilon (bv^{\phi})^2).
  \label{eq:psi_circ2}
\end{eqnarray}

If the observer and source of the light ray are located
in an asymptotically flat region at infinity, $\phi \rightarrow 0$
and $\phi \rightarrow \pi$, respectively, then 
\begin{eqnarray}
 \psi_{\rm transverse}(\phi \rightarrow 0)
  &=&
  \psi_{\rm transverse}(\phi \rightarrow \pi)
  \nonumber\\
  &\rightarrow&
  \left(
   \frac{2m}{b} + \frac{15\pi m^2}{8b^2} - \frac{2am}{b^2}
	 \right)
  \left(1 + \frac{bv^{\phi}}{2}\right)
  \nonumber\\
  &+& {\cal O}(\varepsilon^3, (bv^{\phi})^2, \varepsilon (bv^{\phi})^2).
    \label{eq:psi_circ3}
\end{eqnarray}
Therefore, the total deflection angle $\alpha_{\rm transverse}$
can be obtained as
\begin{eqnarray}
 \alpha_{\rm transverse} &=&
  \psi_{\rm transverse}(\phi \rightarrow 0)
  + \psi_{\rm transverse}(\phi \rightarrow \pi)
  \nonumber\\
  &=&
  \left(
   \frac{4m}{b} + \frac{15\pi m^2}{4b^2} - \frac{4am}{b^2}
  \right)
  \left(1 + \frac{bv^{\phi}}{2}\right)
  + {\cal O}(\varepsilon^3, (bv^{\phi})^2, \varepsilon (bv^{\phi})^2).
 \label{eq:alpha_circ}
\end{eqnarray}
From Eq. (\ref{eq:alpha_circ}) we find that when the observer moves
counterclockwise with respect to the radial direction, $v^{\phi} > 0$,
the measured angle $\psi_{\rm transverse}$ becomes larger than that of
the static observer, and vice versa.

From Eqs. (\ref{eq:alpah_sta}) and (\ref{eq:alpha_circ}),
the relation between $\alpha_{\rm static}$ and
$\alpha_{\rm transverse}$ becomes
\begin{eqnarray}
 \alpha_{\rm transverse}
  = \left(1 + \frac{bv^{\phi}}{2}\right)\alpha_{\rm static}.
  \label{eq:alpha_sta_tra}
\end{eqnarray}
It seems that the form of the overall scaling factor is characterized by
half of the transverse velocity $bv^{\phi}/2$ unlike $v^r$
in the radial motion case; see Eq. (\ref{eq:alpha_sta_rad}).
However, this may depend on the choice of form of the transverse
velocity.
\subsection{Observability in Current and Future Solar System Measurements
\label{sec:observation}}
In this section, let us investigate the observability of the contribution 
of the motion of the observer due to the radial velocity $v^r$ and 
the transverse velocity $bv^{\phi}$ as well as the spin parameter $a$  
on the measurable angle $\psi_P$. As an actual observations 
in the solar system, we discuss observability on the basis of 
the accuracy of VLBI, GAIA, and LATOR, namely $0.1 \sim 10$ mas 
or more better (VLBI), $1 \sim 100\ \mu$as (GAIA), 
and $0.01 ~{\rm picorad} = 10^{-14}~{\rm rad} \simeq 2.0 ~{\rm nas}$ (LATOR), 
respectively. To estimate concrete values below, we adopt the Sun as 
the central (lens) object and take the third escape velocity with respect 
to the Sun as the radial velocity and the transverse velocity of the observer,
$v^r = bv^{\phi} = v_3/c \sim 1.4 \times 10^{-4}$. 
TABLE \ref{tab:arakida-table1} lists the numerical values such as
the mass of the sum $m$, the spin parameter $a$, and impact parameter $b$.
\begin{table}[b]
 \tbl{\label{tab:arakida-table1} Numerical values.
 We use the following numerical values in this paper.
 As the value of the angular momentum of the Sun, we adopt
 $J_{\odot} \simeq 2.0 \times 10^{41}$ kgm$^2$/s from
 \cite{komm_etal2003,bi_etal2011}.}{
\begin{tabular}{ccr}
\toprule
\textrm{Name}&
\textrm{Symbol}&
\textrm{Value}\\
\colrule
 Mass of the Sun & $m = GM_{\odot}/c^2$ & $1.5 \times 10^4$ m\\
 Impact Parameter & $b = R_{\odot}$ & $1.7 \times 10^8$ m\\
 Angular Momentum of the Sun \cite{komm_etal2003,bi_etal2011}
 & $J_{\odot}$ & $2.0\times 10^{41}$ kgm$^2$/s\\
 Spin Parameter & $a = J_{\odot}/(M_{\odot}c)$ & $333.3$ m\\ 
 Third Escape Velocity w.r.t the Sun & $v_3$ & $4.2 \times 10^4$ m/s \\
 Radial Velocity & $v^r = v_3/c$ & $1.4 \times 10^{-4}$\\
 Transverse Velocity & $bv^{\phi} = v_3/c$ & $1.4 \times 10^{-4}$\\
\botrule
\end{tabular}
}
\end{table}
Before estimating the contribution of the velocity of 
the observer $v^r$ and $bv^{\phi}$, we summarize the values of the
measurable angle in the static observer;
\begin{align}
 \frac{2m}{b} \approx 0''.875,\quad
 \frac{15\pi m^2}{8b^2} \approx 5.5\ \mu{\rm as},\quad
 \frac{2am}{b^2} \approx 0.35\ {\rm nas}. 
\end{align}
Therefore, second post-Newtonian term $15\pi m^2/8b^2$ is 
currently detectable by the GAIA. However, in order to observe 
the effect of the rotation of celestial object $2am/b^2$, 
the accuracy of next-generation observation such as LATOR is required.

\begin{figure}[htbp]
\begin{center}
 \includegraphics[scale=0.6,clip]{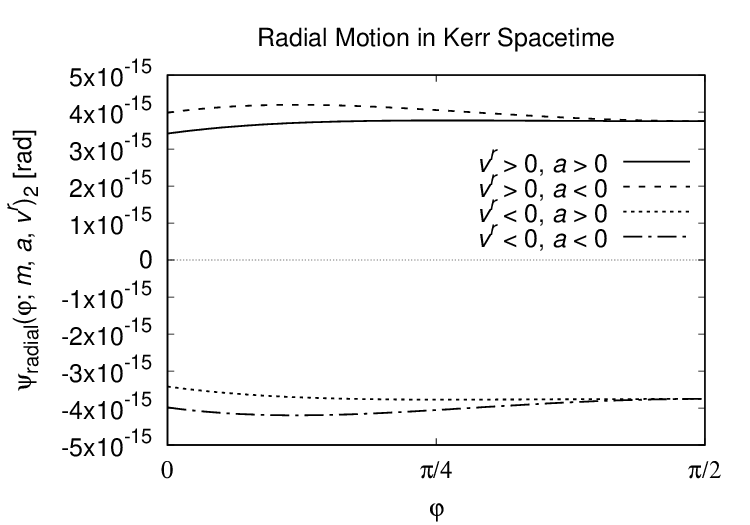}
 \caption{$\phi$ dependence of Eq. (\ref{eq:psi_rad_v2}).
 \label{fig:arakida-fig2}}
\end{center}
\end{figure}

Next, let us deal with the case of the radially moving observer with 
velocity $v^r$. We assume that the observer is located within the range
$0 \le \phi \le \pi/2$, and we extract the terms pertaining to $v^r$
from Eq. (\ref{eq:psi_rad1}):
\begin{eqnarray}
 \psi_{\rm radial}(\phi; m, a, v^r)_0
  &=& v^r \sin\phi,
  \label{eq:psi_rad_v0}\\
 \psi_{\rm radial}(\phi; m, a, v^r)_1
  &=& \frac{2mv^r}{b},
  \label{eq:psi_rad_v1}\\
 \psi_{\rm radial}(\phi; m, a, v^r)_2
  &=&
 \frac{v^r}{16b^2}
  \left\{
	m^2\left[
      30 (\pi - 2\phi) \cos\phi + 95\sin\phi - \sin 3\phi
     \right]\right.
	\nonumber\\
    &-&\left.16am(1 + \cos 2\phi) - 2a^2(3\sin\phi - \sin 3\phi)\right\}.
     \label{eq:psi_rad_v2}
\end{eqnarray}
Eq. (\ref{eq:psi_rad_v0}) can be regarded as the usual aberration term
in the background (flat) spacetime. Then, this term Eq. (\ref{eq:psi_rad_v0}) 
is canceled by taking the difference from the angle $\psi_P^{\rm BG}$
in the background spacetime (BG means the background).
Eq. (\ref{eq:psi_rad_v1}) has a constant value 
$2mv^r/b \simeq 2.4 \times 10^{-10} \simeq 50\ \mu{\rm as}$ throughout 
the range $0 \le \phi \le \pi/2$. 
This value is at the observation limit of VLBI, but it can 
be observed by GAIA.
Figure \ref{fig:arakida-fig2} shows Eq. (\ref{eq:psi_rad_v2})
as a function of $\phi$.
Eq. (\ref{eq:psi_rad_v2}) can be considered as the 2.5 post-Newtonian 
contribution, but the order ${\cal O}(10^{-15}) \simeq 0.2\ {\rm nas}$ 
is almost one order of magnitude smaller than the observational 
limit of LATOR, ${\cal O}(10^{-14}) \simeq 2.0\ {\rm nas}$.
However, with improvements in Interferometry technology and miniaturization 
of optical lattice clock, 
this observational limitation, the value $0.2\ {\rm nas}$, 
may be overcome by the future space mission.

Lastly, we examine the influence of the transverse motion of the observer.
Just as was done for radial motion, from Eq. (\ref{eq:psi_circ1}) we
extract the terms concerning $v^{\phi}$ for the range $0 \le \phi \le
\pi/2$:
\begin{eqnarray}
 \psi_{\rm transverse}(\phi; m, a, bv^{\phi})_0
  &=& bv^{\phi}\tan\frac{\phi}{2},
  \label{eq:psi_circ_v0}\\
 \psi_{\rm transverse}(\phi; m, a, bv^{\phi})_1
  &=& \frac{2mbv^{\phi}\cos\phi}{b(1 + \cos\phi)},
  \label{eq:psi_circ_v1}\\
  \psi_{\rm transverse}(\phi; m, a, bv^{\phi})_2
  &=&
  \frac{bv^{\phi}}{8b^2(1 + \cos\phi)}
  \nonumber\\
 &\times&
  \biggl\{
  m^2\left[15(\pi - 2\phi)
      - 16\sin \phi + 7\sin 2\phi + 16\tan \frac{\phi}{2}
     \right]\biggr.
  \nonumber\\
 &+&
  \biggl.
  8ma(1 - 2\cos\phi - \cos 2\phi)
  \biggr\}.
  \label{eq:psi_circ_v2}
\end{eqnarray}
We choose the transverse velocity in the same way that we chose radial 
motion: $bv^{\phi} = v_3/c = 1.4 \times 10^{-4}$.

\begin{figure}[htbp]
\begin{center}
 \includegraphics[scale=0.6,clip]{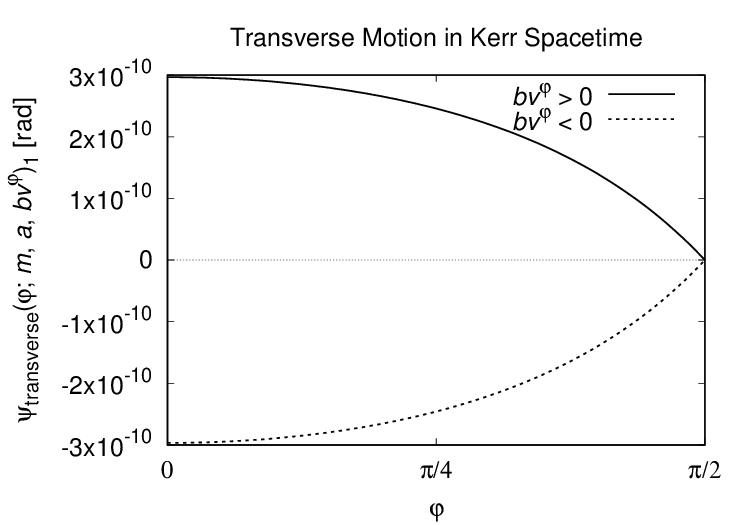}
 \caption{$\phi$ dependence of
 Eq. (\ref{eq:psi_circ_v1}).
 \label{fig:arakida-fig3}}
\end{center}
\end{figure}

\begin{figure}[htbp]
\begin{center}
 \includegraphics[scale=0.6,clip]{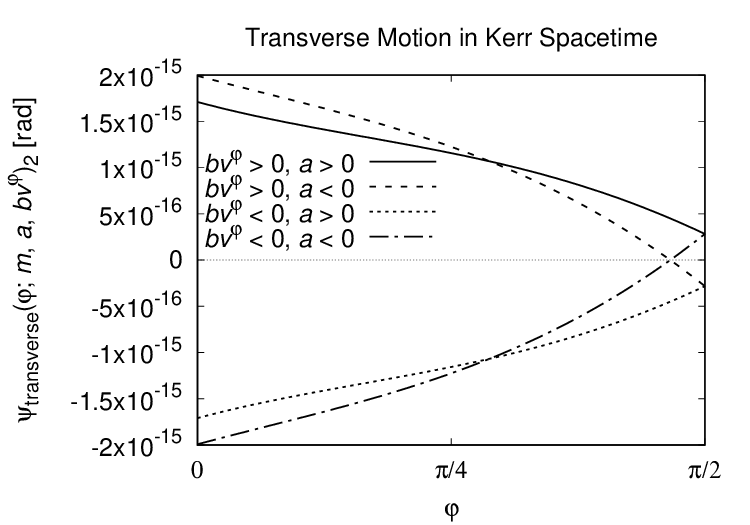}
 \caption{$\phi$ dependence of
 Eq. (\ref{eq:psi_circ_v2}).
 \label{fig:arakida-fig4}}
\end{center}
\end{figure}

Eq. (\ref{eq:psi_circ_v0}) is canceled by taking the difference from 
the aberration term in the background spacetime for the same reason 
as Eq. (\ref{eq:psi_rad_v0}). 
FIG. \ref{fig:arakida-fig3} plots Eq. (\ref{eq:psi_circ_v1}) as 
a function of $\phi$ which is almost order 
${\cal O}(10^{-10}) \simeq 20\ \mu{\rm as}$.
Therefore the value of Eq. (\ref{eq:psi_circ_v1}) is at the observation 
limit of VLBI, and lies in the observable range of GAIA.
FIG. \ref{fig:arakida-fig4} illustrates the $\phi$ dependence of
Eq. (\ref{eq:psi_circ_v2}).
As for Eq. (\ref{eq:psi_rad_v2}), Eq. (\ref{eq:psi_circ_v2}) 
corresponds to the 2.5 post-Newtonian contribution. From FIG. \ref{fig:arakida-fig4}, the value of
Eq. (\ref{eq:psi_circ_v2}) is the order 
${\cal O}(10^{-15}) \simeq 0.2\ {\rm nas}$, 
however it is almost one order of 
magnitude smaller than the observational threshold, e.g., 
${\cal O}(10^{-14}) \simeq 2.0\ {\rm nas}$ of LATOR. 
Nonetheless this value $0.2\ {\rm nas}$ can be expected to be 
observable in future mission.

In summary, we list the estimated values of the measurable angle 
in this section in table \ref{tab:arakida-table2}.
\begin{table}[b]
 \tbl{\label{tab:arakida-table2} Estimated values of measurable angle and
 observability in current and future measurements, VLBI, GAIA, and LATOR.
 The following symbols mean; $\bigcirc$ is the observable, 
 $\triangle$ is the limit of observation, and $\times$ is currently not
 observable. 1pN, 2pN, 1.5pN, and 2.5pN indicate the order of the
 post-Newtonian approximation.}{
\begin{tabular}{lcrccc}
\toprule
\textrm{Term}&
\textrm{Order of Approximation}&
\textrm{Estimated Value}&
\textrm{VLBI}&
\textrm{GAIA}&
\textrm{LATOR}\\
\colrule
\textrm{Static parts} &      &             &          &          &        \\ 
\colrule
 $2m/b$  & 
 1pN & 
 $0''.875$   & 
 $\bigcirc$  & 
 $\bigcirc$  & 
 $\bigcirc$\\
 $15\pi m^2/8b^2$ 
 & 2pN & 
 $5.5\ \mu {\rm as}$  &
  $\times$ & 
  $\bigcirc$  &
   $\bigcirc$\\
 $2am/b^2$    & 
 2pN & 
 $0.35\ {\rm nas}$ & 
 $\times$ & 
 $\times$ & 
 $\times$\\
\colrule
\textrm{Radial parts} &        &        &   &                    &        \\ 
\colrule
 Eq. (\ref{eq:psi_rad_v1}) & 
 1.5pN & 
 $50\ \mu {\rm as}$ &
 $\triangle$ & 
 $\bigcirc$ &
 $\bigcirc$\\ 
 Eq. (\ref{eq:psi_rad_v2}) & 
 2.5pN & 
 $0.2\ {\rm nas}$   & 
 $\times$    & 
 $\times$ & 
 $\times$ \\
\colrule
\textrm{Transverse parts} &   &    &            &              &        \\ 
\colrule
 Eq. (\ref{eq:psi_circ_v1}) & 
 1.5pN & 
 $20\ \mu {\rm as}$ & 
 $\triangle$ & 
 $\bigcirc$ & 
 $\bigcirc$\\
 Eq. (\ref{eq:psi_circ_v2}) & 
 2.5pN & 
 $0.2\ {\rm nas}$    & 
 $\times$    & 
 $\times$ & 
 $\times$\\
\botrule
\end{tabular}
}
\end{table}
\section{Conclusions\label{sec:conclusions}}
We focused on the measurable angle (local angle) of the light ray
$\psi_P$ at the position of the observer $P$ instead of the total 
deflection angle (global angle) $\alpha$ in Kerr spacetime.
We investigated not only the effect of frame dragging represented
by the spin parameter $a$ but also the influence of the motion of
the observer with the coordinate radial velocity $v^r$ and
the coordinate transverse velocity $bv^{\phi}$ ($v^{\phi}$ is the 
coordinate angular velocity), which are related to the 4-velocity
of the observer $u^r$ and $u^{\phi}$, respectively.

To consider the influence of the velocity of the observer
$v^r$ and $bv^{\phi}$ on the measurable angle $\psi$,
we employed the general relativistic aberration equation
because the motion of the observer changes the direction of
the light ray approaching the observer. 
The local measurable angle $\psi_P$ obtained in this paper can be
applied not only to the case of the observer located in
an asymptotically flat region but also to the case of the observer
placed within the curved and finite-distance region.

When the observer is moving in the radial direction, the total
deflection angle $\alpha_{\rm radial}$ can be expressed by
$\alpha_{\rm radial} = (1 + v^r)\alpha_{\rm static}$ which is
equivalent to the results obtained by, e.g., 
\cite{pyne_birkinshaw1993,frittelli2003,wucknitz_sperhake2004},
whose overall scaling factor is $1 - v$ instead of $1 - 2v$
obtained by, e.g., \cite{capozziello_etal1999,sereno2002}.
We notice that $v^r$ and $v$ are related with $v^r = -v$
because the velocity $v$ in
\cite{pyne_birkinshaw1993,capozziello_etal1999,sereno2002,frittelli2003,wucknitz_sperhake2004}
is that of the lens object. On the other hand, when the observer
moves in the transverse direction, the total deflection angle is
given by the form
$\alpha_{\rm transverse} = (1 + bv^{\phi}/2)\alpha_{\rm static}$
if we define the transverse velocity as having the form of $bv^{\phi}$.
However, this relation may depend on the choice of transverse
velocity.

As we observed, the order ${\cal O}(mv^r)$ and ${\cal O}(mbv^{\phi})$
terms in Eqs (\ref{eq:psi_rad_v1}) and (\ref{eq:psi_circ_v1})
are of order $10^{-10}$, and therefore they are at the observational 
limit of the VLBI but they are in the observable range of planned space 
missions such as LATOR. However, the order ${\cal O}(m^2v^r, a^2v^r, mav^r)$ and
${\cal O}(m^2bv^{\phi}, a^2bv^{\phi}, mabv^{\phi})$ terms in
Eqs (\ref{eq:psi_rad_v2}) and (\ref{eq:psi_circ_v2}) are of order
$10^{-15}$, and thus they are one order of magnitude smaller than
the observation limit, e.g., $10^{-14}$, of LATOR.
Nevertheless, this limitation may be overcome in the near future.

\section*{Acknowledgments}
We would like to acknowledge anonymous referees for reading our manuscript 
carefully and for giving fruitful comments and suggestions, which significantly 
improved the quality of the manuscript. 


\end{document}